# Chemical Physics Letters

# Visualisation of single atom dynamics and their role in nanocatalysts under controlled reaction environments

Pratibha L. Gai [a,b,⇑], Leonardo Lari [b], Michael R. Ward [b], Edward D. Boyes [b,c]

[a] The Nanocentre, Department of Chemistry, University of York, York YO10 5DD, UK
[b] The Nanocentre, Department of Physics, University of York, York YO10 5DD, UK [c] The Nanocentre, Department of Electronics, University of York, York YO10 5DD, UK

## ABSTRACT

Direct real time studies of reacting individual atoms in technologically important gold and platinum nanocatalysts in controlled reducing and oxidising gas environments and operating temperatures using novel environmental scanning transmission electron microscopy (ESTEM) with single atom sensitivity are presented. The direct in situ observations provide new insights into the dynamic behaviour of single atoms which may be important as catalytic active sites as well as migrating as part of deactivation mechanisms. The single atom dynamics reveal that the primary role of nanoparticles in the catalysts is to act as reservoir of ad-atoms or clusters. Possible reaction mechanisms are described briefly.

## 1. Introduction

Supported gold and platinum nanoparticle catalysts play a pivotal role in a wide range of chemical reactions in catalysis in the production of chemicals, energy, and for environmental emission control. For example, Au nanoparticle catalysts are of interest in hydrogenation, water–gas-shift reactions and low temperature oxidation of carbon monoxide [1] and platinum nanocatalysts play a key role in the chemical industry including in emission control and fuel cells [2]. There is increasing indirect indication that single atoms or clusters of a few atoms may play a key role in catalytic reactions [3,4]. However the in situ observation of reacting single atoms (about 100 picometers in size) under severe reaction conditions of gas environments and operating temperatures is extremely difficult. Direct visualisation of reacting single atom dynamics is critical to obtaining insights into mechanisms of chemical reactions important in the development of novel materials and processes.

The electron microscope (EM) is a powerful tool to study catalyst nanostructures but conventional EM operates in high vacuum. Previously Gai and Boyes reported the world's first development of the atomic resolution environmental transmission electron microscope (atomic resolution ETEM) which enabled real time atomic resolution observations of gas–solid reactions under controlled conditions of a flowing gas environment around the sample (as in technological reactors) with pressures up to 1000s of Pa and at the operating reaction temperature and for in situ nanosynthesis [5,6]. Highlights of their development included the novel ETEM design with objective lens polepiece incorporating radial holes for differential pumping and the regular EM sample chamber used as the controlled reaction environmental cell or reactor [6]. The ETEM development paved the way to atomic resolution studies of gas (or liquid)–solid reactions at operating temperatures under controlled conditions allowing observations of reacting columns of atoms and is now widely used worldwide [e.g., [7–9]].

## 2. Materials and Methods

Recently the development of the first double aberration corrected E(S)TEM (AC E(S)TEM) utilising a large gap objective lens polepiece with the sample under controlled reaction conditions of a flowing gas atmosphere and high temperatures (>500 °C) for directly visualising single atoms and clusters has been reported by Boyes and Gai [10–12]. These studies include calibration procedures to remove any deleterious effects of the electron beam on the sample which are described in references [5–12].

In this Letter we present in situ real time studies of dynamic Au and Pt nanocatalyst reactions at the single atom level in controlled reducing and oxidising reactive flowing gas atmospheres and temperature using AC ESTEM, and demonstrate single atom dynamics and their role in catalyst structural changes and reaction mechanisms. With our new AC-ESTEM capability it is now possible to analyse real space STEM Z-(atomic number) contrast images under transient conditions with full analytical functionalities (including chemical composition, diffraction and spectroscopy). The images in this Letter are all high angle annular dark field (HAADF)-STEM images using a JEOL 2200FS (S)TEM instrument which we have

⇑ Corresponding author at: The Nanocentre, Department of Chemistry, University of York, York YO10 5DD, UK.
E-mail addresses: pratibha.gai@york.ac.uk (P.L. Gai), ed.boyes@york.ac.uk (E.D. Boyes).

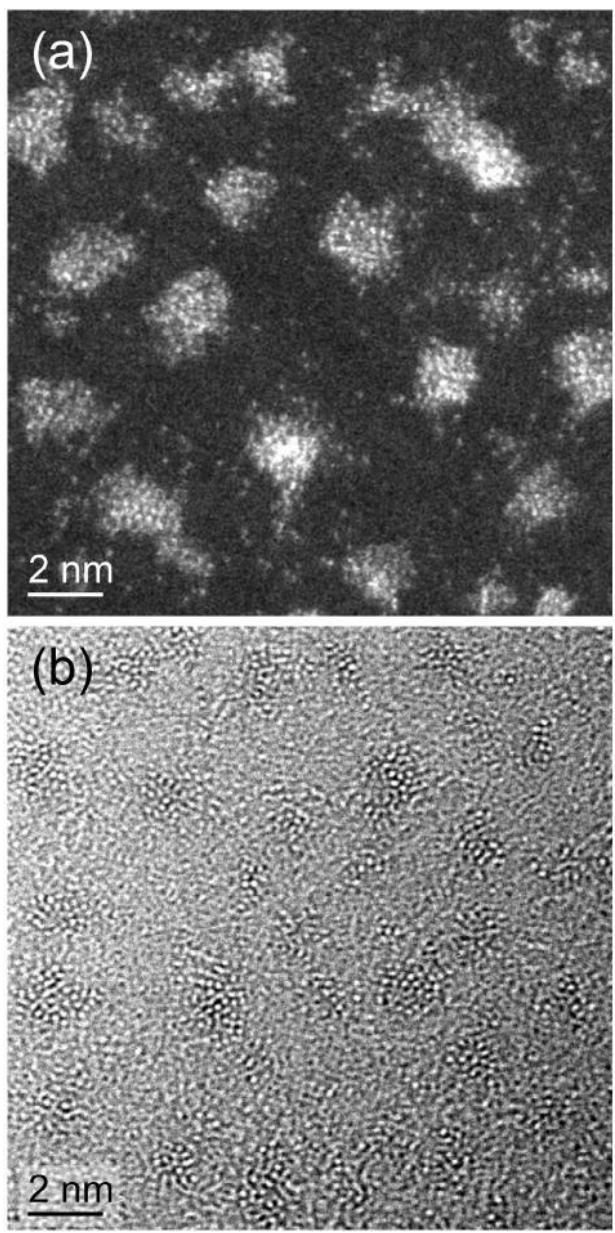

Figure 1. Single atoms and clusters/nanoparticles of Pt/C catalyst (a) AC ESTEM HAADF; (b) in AC ETEM.

modified in-house at York [10–12]. In HAADF electrons that undergo Rutherford scattering are collected and the intensity is approximately proportional to $Z^2$ (Crewe et al.) [13]. The ESTEM calibration procedures for imaging and recording are reported earlier [10–12]. Briefly magnifications of 8-12MX with 512 and 1024 line frames and sampling at 0.015 nm and 0.03 nm were used. For controlled heating experiments in flowing gas a Gatan 628 hot stage was used with custom electronics and the model catalysts were prepared by sputtering the metals on carbon supports.

3. Results and Discussion

HAADF AC ESTEM is preferable for single atom analysis compared to AC ETEM as shown in the following. An ESTEM image of single atoms and clusters in a Pt/C model catalyst is shown in Figure 1a. In comparison, in the ETEM image shown in b diffraction effects from the support contrast make it difficult to discern single atoms clearly and impossible to analyse them quantitatively.

Figure 2a shows an ESTEM HAADF image of single atoms, clusters and nanoparticles of as prepared gold nanocatalysts, recorded in vacuum. The intensity profile of a single atom is shown in Figure 2b. Single atom images measure about 0.11 nm in diameter. Figure 2c summarises preliminary data of plots of ESTEM HAADF intensities from the unheated support film (indicated by A) showing the profiles of single atoms (B) and clusters and nanoparticles (C) on the support of Au/C sample. The image in Figure 2a shows a combination of single atoms (indicated by a white rectangle on the image), flatter and more raft like clusters of atoms, and more developed nanoparticles (indicated by an arrow). The original Crewe et al. [13] HAADF paper included quantitative measurements of a restricted number (n) of atoms in each column (up to n < 10 atoms per column). The sensitivity and accuracy of these analyses have now improved using higher resolution aberration corrected EM instruments [10–12,14,15]. In this Letter we identify single atoms in a similar way to Batson et al. [14] and present initial illustrative data in Figure 2c.

In our analyses, the graph in Figure 2c was obtained by taking 3 x 3 pixel squares and manually sampling the background, single atoms and the clusters in that order from the same image shown in Figure 2a. About 120 samples were taken from the background and nanoparticles and about 60 single atoms (isolated from the nanoparticles or clusters) were selected. An average intensity was deduced from each square and the error bars are standard deviations. The initial data show discrete levels in the video signals from single atoms; and multiple (n up to about 4) 'layers' of atoms in the initial clusters; based on a linear model [13]. Single atoms, clusters, and nanoparticles in Figure 2a have been analysed by video intensity of discrete columns of atoms in a regular crystal structure (Figure 2c). A detailed analysis of atom column intensities in the selected nanoparticle from Figure 2c show particle edges made up of minimum columns containing multiple atoms; compared to the initial rafts with greater area and tapering to single atom high edges. To reduce the effects of signal auto-levelling in the electronics, the video signals are compared only within each frame where single atoms can be used for reference to calibrate the local video signal following the normalisation of data between frames. The error bars reflect noise in the signals of the individual measurements. The analyses of the column intensities support the observations.

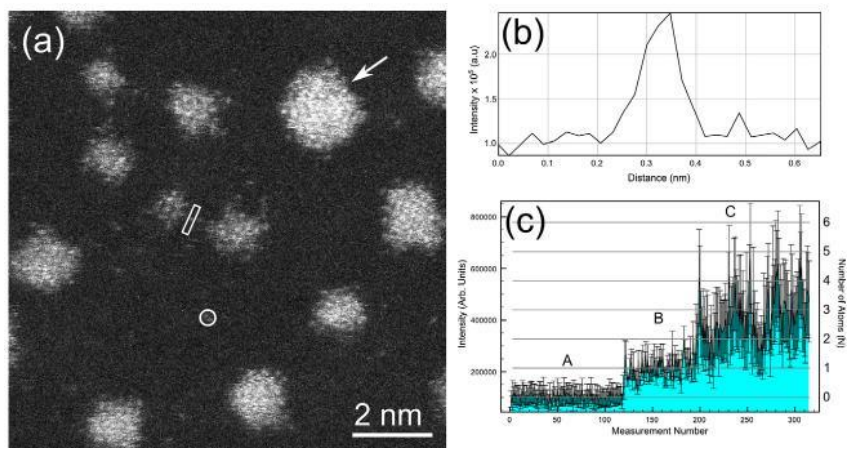

Figure 2. (a) Au/C catalyst with single atoms and clusters (b) intensity profile of the single atom image indicated; (c) plots of ESTEM HAADF intensities of Au/C nanocatalyst at individual atomic (column) positions (C) normalised to single gold atom with the average background signal (B–A) removed so that for small column heights (N (number) of atoms) = C/(B–A). A: area of support film; B: as deposited single atoms; C: clusters and nanoparticles.

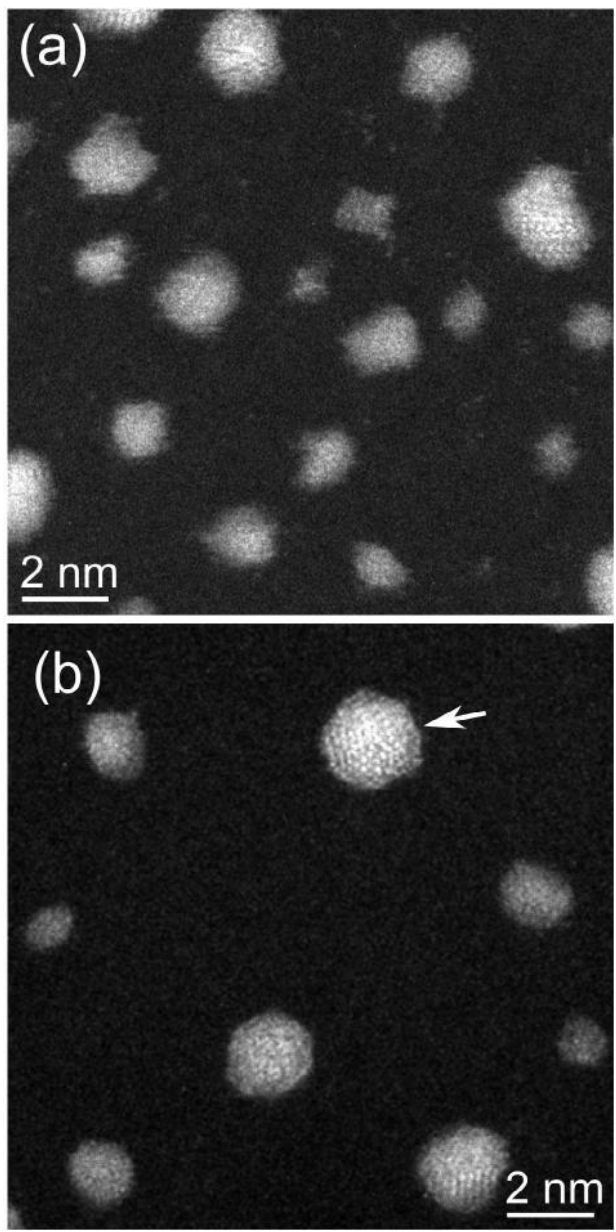

Figure 3. Au/C catalyst reaction in flowing Hydrogen gas at the single atom level and at operating temperatures: (a) Au/C at RT (b) in hydrogen gas at 500 °C indicating increased faceting of particles. A more faceted particle is shown by an arrow in (b).

Figure 3a and b shows Au nanoparticles on carbon at room temperature (RT) and in flowing hydrogen gas environment of 200 Pa inlet pressure at 500 °C. We have measured that the gas pressure at the sample is about 2 Pa. The data at the single atom level reveal the onset of more faceted particles (e.g., indicated by an arrow) and fewer single atoms in the gas environment at the operating temperature. The results presented here show the dynamic behaviour of single atoms. We believe that in hydrogen gas low coordination surface atoms on nanoparticles are replaced by facets to minimise the surface energy of the particle.

Further direct evidence of single atom dynamics and increased faceting in Pt/C catalysts is shown in Figure 4, consistent with our observations in the $H_2$ reduction of gold nanocatalysts. It illustrates a sequence of a small particle (P) undergoing structural rearrangement. Figure 4b forms the intermediate stage, indicating the migration of single atoms from the particles (appearing as white dots (arrowed) on the thin amorphous carbon support film, leading to new clusters and more faceted particles. The driving force for the removal of atoms away from the particle is the minimisation of the particle surface energy, where surface ad-atoms and steps are removed in order to create lower energy facets. The onset of faceting can be seen in Figure 4c and the formation of small clusters.

For reduction–oxidation studies at the single atom level of Pt/C nanocatalysts which are important in fuel cell applications, the reduction was first carried out in flowing hydrogen gas with the inlet gas pressure of 190 Pa at 500 °C. The sample was then cooled to about 200 °C without the gas and oxygen gas was introduced with the inlet gas pressure of 120 Pa. Figure 5 shows reduction–oxidation sequence of the nanocatalyst: (a) the catalyst in vacuum; (b) in hydrogen at 500 cC and (c) the sample in oxygen gas at 200 °C. In Figure 5b there appear to be fewer single Pt atoms at this temperature, consistent with our observations in the $H_2$ reduction of gold nanocatalysts. The observations suggest that the migration of loosely bound single atoms contribute to particle shape changes. More single atoms appear to be present in oxygen gas (than in hydrogen) as illustrated in Figure 5c. This is discussed in the following sections.

In-situ hydrogen reduction in Figure 5b shows the presence of fewer single atoms on the support than in a, clusters and increased faceting of the nanoparticles. This indicates that the nanoparticles act as a source of potential ad-atoms and clusters with important implications in catalysis and understanding the role of nanoparticles. Low coordination surface atoms are believed to enhance catalytic activity of small nanoparticles [16,17]. The minimisation of surface energy leads to the removal of low coordination atoms and stable surface facets of the nanoparticles. This decrease in the low coordination atoms is expected to lead to the deactivation of nanoparticles and may explain the often observed initial drop in the activity of nanocatalysts. Through the use of appropriate supports with anchoring sites [18,19] such migratory single atoms may be anchored and stabilised which will assist in providing active sites for the gas reaction. In oxygen relatively more single atoms appear to be present than in hydrogen at higher temperatures as shown in Figure 5c (than in b), indicating that the single atom reaction with oxygen at lower temperature prevents their migration. This is discussed below.

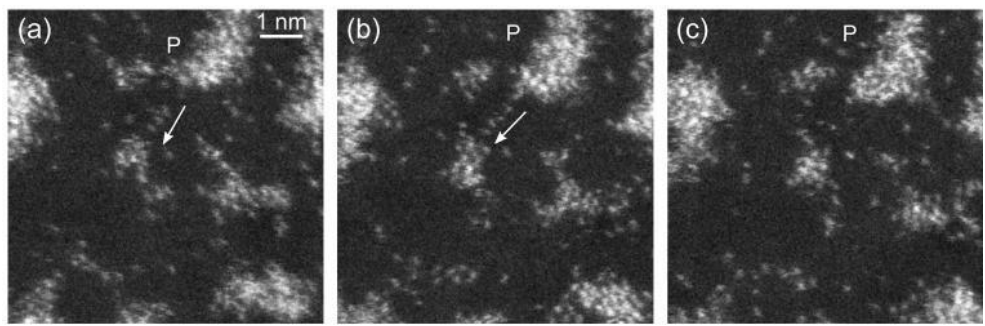

Figure 4. Single atom dynamics in Pt/C catalysts: migration of single atoms arrowed in (a) and (b) (e.g., from a cluster/particle P) leading to increased faceting of the particle and the formation of clusters (c). (ESTEM).

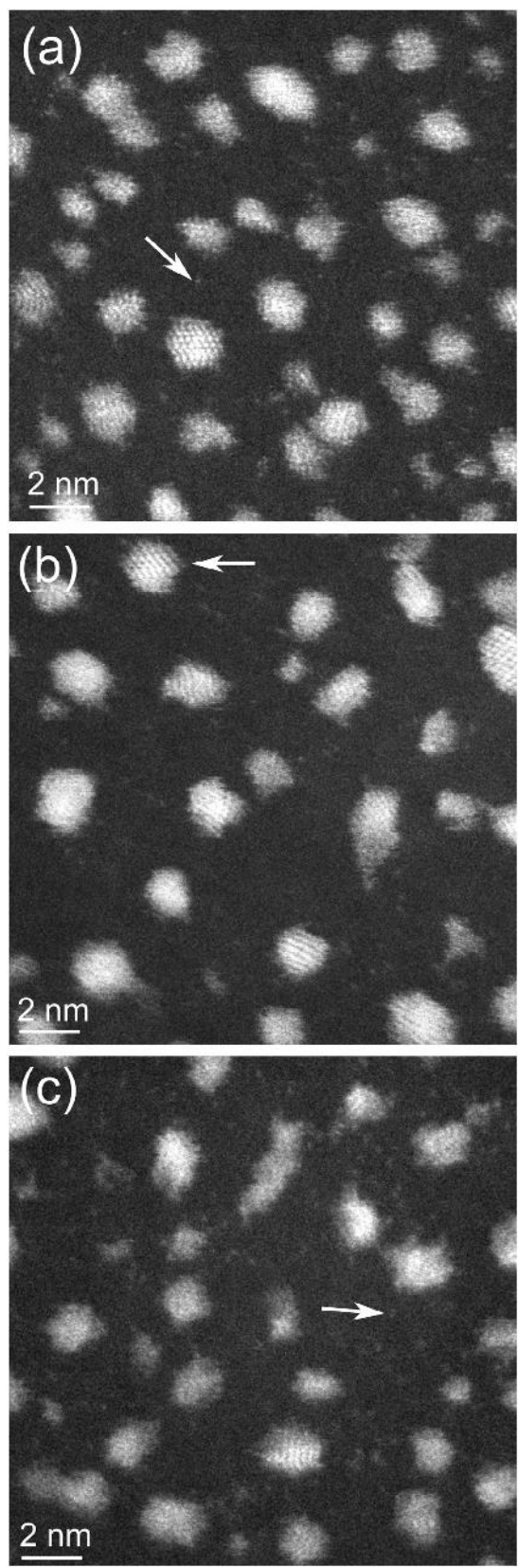

Figure 5. Reduction–oxidation reactions at the single atom level in ESTEM. Reacting single atoms in Pt/C nanocatalyst in reactive gas environments. The mobility of single atoms is shown to be temperature dependent: (a) at RT. Single atom is indicated by an arrow; (b) flowing hydrogen gas at 500 °C; showing faceting and fewer single atoms at the higher temperature; A more faceted particle is indicated by an arrow (c) flowing oxygen gas at 200 °C with single atoms (e.g. arrowed) and interaction with $O_2$.

Although the behaviour of Pt single atoms in oxygen gas at elevated temperatures is complex, some tentative conclusions can be drawn from our in situ studies of reacting individual atoms under controlled reaction conditions. Our initial data enabled by our new ESTEM capability suggests that in the presence of hydrogen there is morphological and nanostructural evolution of an initially finely divided Pt model system on a carbon support film. Perhaps the most surprising discovery is the wide distribution of single Pt atoms individually dispersed; and revealed for the first time with our new ESTEM HAADF Z-contrast imaging capability. As well as their established role in competitive particle growth through an Ostwald ripening mechanism, the residence of individual atoms on the support may be expected to have energy states and attendant catalytic activity and specific selectivity quite different from the properties of Pt atoms with other Pt atoms as nearest neighbours. They can be thought of, perhaps, as extreme kink sites. And very different to the more bulk-like properties of larger entities, in this case >2 nm in size, with essentially a crystalline bulk Pt internal structure and corresponding surfaces with relatively few low co-ordination sites. With the AC ESTEM it is now possible with considerable precision to plot the position of every atom in a nanoparticle up to perhaps 200 atoms in population. As well as single atoms partially ordered 'rafts' of Pt atoms develop. These are typically only 1 atom high at the edges and generally no more than 2 high in the centre. It appears they are in some dynamic equilibrium with the population of dispersed single atoms.

The mobility of single atoms appears to be temperature dependent. After heating at 500 °C in a flowing $H_2$ environment the number of single atoms detected is greatly reduced. We believe that this may be due to, either because of detectability issues with short site specific residence times which may be modified by the action of the gas or representing a changed state, or some combination and these and perhaps other factors which are presently unclear. With prolonged treatment in $H_2$ at 500 °C more faceted cuboid single crystals of Pt are produced (e.g., shown by arrow in Figure 5b). They seem to have 'cliff edge' {200} side (and top) faces a minimum of about 4 atoms high in the observations and to have evolved from the preliminary rafts; themselves assembled from the initial population of single atoms (as observed in Figure 4). After the specimen is cooled down to room temperature (RT) and the $H_2$ removed, more single atoms are again detected and this effect appears to be present when the specimen is treated at 200 °C in a flowing oxygen environment (at much lower temperature than in hydrogen gas in (b)). These initial observations are illustrated in Figure 5 and they suggest that the understanding of the role of single atoms in heterogeneous catalysis may have to be extended considerably.

At 200 °C, our data suggest interactions between Pt atoms and the chemisorbed oxygen, as well as interactions with nanoparticles which exhibit irregular structures and less faceting (Figure 5c). There are reports with gold that particle shape changes can occur at ambient temperatures [20]. Based on the observation in Figure 5c we believe that in $O_2$ environments Pt binds oxygen strongly, preventing the migration of low coordination atoms, indicating the interaction of oxygen on low co-ordination atoms. We believe that the binding is critical to the efficacy of Pt as catalysts especially in fuel cells, where binding with $O_2$ and reaction with $H_2$ are important. These observations in hydrogen and oxygen indicate that Pt atom dynamics are likely to be temperature dependent in oxygen and single atoms are important in increasing the effective Pt surface area and the intrinsic activity per site.

## 4. Conclusions

In summary, the newly discovered dynamic single atom interactions and with nanoparticles in controlled reduction–oxidation reactions demonstrate their function in the chemical reactions with the primary role of nanoparticles in the reactions studied being to act as reservoir or source of ad-atoms and clusters. The results have important implications in catalysis and studies of nano-particles generally and a radical revision of the interpretation of the activity, loss of surface area due to sintering and deactivation of catalysts may be required.

## Acknowledgements

We thank the EPSRC (UK) for critical mass research grant EP/J018058/1. Ian Wright is thanked for support.